\documentclass[12pt]{article}
\usepackage{amsmath}
\usepackage{epsfig}
\def\la{\mathrel{\mathpalette\fun <}}
\def\fun#1#2{\lower3.6pt\vbox{\baselineskip0pt\lineskip.9pt
\ialign{$\mathsurround=0pt#1\hfil##\hfil$\crcr#2\crcr\sim\crcr}}}

\begin{document}
\title{Asymptotic behavior of photoionization cross section in a central field.
Ionization of the $p$ states.}
\author{E. G. Drukarev, A. I. Mikhailov\\
{\em National Research Center "Kurchatov Institute"}\\
{\em B. P. Konstantinov Petersburg Nuclear Physics Institute}\\
{\em Gatchina, St. Petersburg 188300, Russia}}
\date{}
\maketitle

\begin{abstract}
{We continue our studies of the high energy nonrelativistic asymptotics for the photoionization cross section of the systems bound by a central field  $V(r)$. We consider the bound states with the orbital momentum $\ell=1$. We show, that as well as for the $s$ states the asymptotics can be obtained without solving of the wave equations for the bound and outgoing electrons. The asymptotics of the cross sections is expressed in terms of the asymptotics of the Fourier transform $V(p)$ of the field and its derivative $V'(p)$ by employing the Lippmann--Schwinger equation. The shape of the energy dependence of the cross sections is determined by the analytical properties of the potential $V(r)$. The cross sections exhibit power drop with the increase of the photon energy for the potentials $V(r)$ which have singularities on the real axis. They experience exponential drop if $V(r)$ has poles in the complex plane. We trace the energy dependence of the ratios
of the photoionization cross sections for $s$ and $p$ electrons from the states with the same principle quantum number.
We apply the results to the physics of fullerenes.
}
\end{abstract}

\section{Introduction}
In our recent paper \cite{1} we studied the high energy photoionization of $s$ states. We presented a more detailed analysis in \cite{2}, \cite{3}. Now we turn to the photoionization of the single-particle states with the orbital momentum $\ell=1$.

We consider the photon energies $\omega$ which are much larger than the ionization potential $I_B$ and find the leading term of expansion of the cross section $\sigma(\omega)$ in terms of $1/\omega$. We assume that the photon energy is much smaller than the electron rest energy $m_e$ (we employ the system of units with $\hbar=1$, $c=1$). Thus we consider the photon energies limited by the conditions
\begin{equation}
I_B \ll \omega \ll m_e,
\label{1}
\end{equation}
analyzing the high energy nonrelativistic asymptotics.

The photoelectron can be treated in nonrelativistic approximation. Its energy is
\begin{equation}
\varepsilon=\omega-I_B=\frac{p^2}{2m_e},
\label{2}
\end{equation}
with $p$ the photoelectron momentum.
Due to Eq.(\ref{1})
\begin{equation}
p \gg \mu,
\label{3}
\end{equation}
where $\mu=(2mI_B)^{1/2}$ is the characteristic momentum of the bound state. Large momentum $q \approx p \gg \mu$ (the "recoil momentum") is transferred to the recoil ion.

We assume the electrons to be bound by a local central field $V(r)<0$. Large recoil momentum $q$ is transferred to the bound electron or to the photoelectron by the source of the field. We show that the relative role of the two mechanisms depends on the behavior of the Fourier transform of the potential $V(r)$ which is  $\tilde V(p)=\int d^3r V(r)e^{-i{\bf p}{\bf r}}$. We omit the tilde sign below. Employing the velocity form of the photon-electron interaction we find that if for large $p \gg \mu$
\begin{equation}
|V'(p)| \gg -\frac{V(p)}{p},
\label{4}
\end{equation}
with $V'(p)=\partial V(p)/\partial p$, the recoil momentum is transferred mostly by the bound electron. However for
\begin{equation}
|V'(p)| \sim -\frac{V(p)}{p},
\label{5}
\end{equation}
transfer of momentum $q$ by both bound and continuum electrons should be included. For example, in the case of the well potential Eq.(\ref{4})
is true, while Eq.(\ref{5}) is true for the Coulomb and exponential potentials.

The potential $V(p)$ should drop at least as $1/p$ at $p \rightarrow \infty$. Otherwise the corresponding potential in the position space $V(r)$ drops faster than $1/r^2$ causing the "fall to the center"\cite{4}. Thus it can not form a bound state. In Sec. 2,3 we analyze the potentials for which
\begin{equation}
pV(p) \rightarrow 0,
\label{8}
\end{equation}
at $p \rightarrow \infty$. We consider a potential for which $V(p) \sim 1/p$ in Subsec.4.1.

We demonstrate that the asymptotics of the photoionization cross section can be obtained without solving the wave equations for the electrons.
The shape of the cross section is determined by the analytical properties of the potential $V(r)$. It is expressed in terms of the potential $V(p)$ and its derivative $V'(p)$. The cross sections for the potentials with singularities at $r=0$ are directly related to the potential $V(p)$. For the potentials with discontinuities  on the real axis the cross sections are expressed in terms of the jumps experienced by the potential $V(r)$. For the potentials with poles in the complex plane the cross sections
experience exponential drop with the power of the exponent determined by the imaginary part of the pole.

Besides the expressions for the cross sections we trace the energy dependence of the ratios
\begin{equation}
r(\omega)=\frac{\sigma_{ns}(\omega)}{\sigma_{np}(\omega)},
\label{8a}
\end{equation}
of the photoionization cross sections for $s$ and $p$ electrons from the states with the same principle quantum number $n$.

We present the general equations for the amplitude and the cross sections in Sec.2. We consider the particular cases in Sec.3. In Sec.4 we apply the results for analysis of photoionization of fullerenes. We summarize in Sec.5.

\section{Asymptotic forms of the amplitude and the cross section}

The general expression for the differential photoionization cross section can be written as
\begin{equation}
d\sigma=n_e\frac{m_ep}{3\pi}\sum_m|F_m|^2\frac{d\Omega}{4\pi}; \quad p=\sqrt{2m_e\omega},
\label{20a}
\end{equation}
with $\Omega$ --the solid angle of the photoelectron. In Eq.(\ref{20a}) $m=0,\pm 1$ is the projection of the angular momentum $\ell=1$ of the bound electron. The averaging over the directions of the photon polarization should be carried out.

The general expression for the amplitude is
\begin{equation}
F_m=N(\omega)\int\frac{d^3f}{(2\pi)^3}\psi^*_{\bf p}({\bf f})\frac{{\bf e}\cdot {\bf f}}{m_e}\psi_m({\bf f}); \quad N(\omega)=\Big(\frac{4\pi\alpha}{2\omega}\Big)^{1/2},
\label{6a}
\end{equation}
with $\psi_{\bf p}({\bf f})$-the wave function of the photoelectron caring the asymptotic momentum ${\bf p}$.

Consider first the mechanism in which the large recoil momentum is transferred by the bound electron.
The photoelectron can be described by the plane wave i.e.$\psi_{\bf p}({\bf f})=\psi_{\bf p}^{(0)}({\bf f})=(2\pi)^3\delta({\bf f}-{\bf p})$ , and the amplitude is
\begin{equation}
F^a_m=N(\omega)\frac{{\bf e}\cdot {\bf p}}{m_e}\psi_m({\bf p}),
\label{6}
\end{equation}
 with $\psi_m$ the single particle wave function of the bound electron.

 Now we must calculate the function $\psi_m(p)$ for $p \gg \mu$.
 This can be done by employing the Lippmann--Schwinger equation
\begin{equation}
\psi_m=\psi^{(0)}_m+G(\varepsilon_B)V\psi_m.
\label{6aa}
\end{equation}
in momentum space.
Here $\psi^{(0)}_m$ is the wave function at $V=0$, $G$ is the propagator of the free motion, $\varepsilon_B=-I_B<0$-is the electron energy in the bound state. For the bound states $\psi^{(0)}_m=0$. The matrix element of the free propagator is
$$\langle {\bf p}|G(\varepsilon_B )|{\bf p}_1\rangle=g(\varepsilon_B, p)\delta({\bf p}-{\bf p}_1); \quad g(\varepsilon_B, p)=\frac{1}{\varepsilon_B-p^2/2m_e}.$$
This provides
$\psi_m({\bf p})=\langle {\bf p}|GV|\psi_m\rangle=g(\varepsilon_B, p)J_m({\bf p}),$
with
\begin{equation}
J_m({\bf p})=\int \frac{d^3f}{(2\pi)^3}V({\bf p}-{\bf f})\psi_m({\bf f}).
\label{6b}
\end{equation}
Hence for any ${\bf p}$
\begin{equation}
\psi_m({\bf p})=g(\varepsilon_B, p)J_m({\bf p}).
\label{6c}
\end{equation}
In the asymptotics $g(\varepsilon_B, p)=-2m_e/p^2$, and thus
\begin{equation}
\psi_m({\bf p})=-\frac{1}{\omega}J_m({\bf p}).
\label{6d}
\end{equation}

Now we analyze the contributions of the three regions of the values of momentum $f$ to the integral $J_m({\bf p})$. The region $f \sim \mu \ll p$, where can estimate $V({\bf p}-{\bf f})=V({\bf p})$
provides a contribution of the order $V(p)/p^2$ to the function $\psi_m({\bf p})$. The  large momenta $f \sim p$ for which $|{\bf p}-{\bf f}| \ll p$  provide the contribution of the order $\psi_m(p)/p^2$ to the right hand side of Eq.(\ref{6d}). It is beyond the asymptotics.
As to the region of large momenta $f \sim p$ ; $|{\bf p}-{\bf f}| \sim p$,
its contribution to the right-hand side of Eq.(\ref{6d})) is of the order $pV(p)\psi_m({\bf p})$. It is also beyond the asymptotics due to Eq.(\ref{8}).

Hence the integral $J_m({\bf p})$ is determined by the region of small momenta  $f \sim \mu \ll p$ for the potentials $V(p)$, which drop faster than $1/p$ at large $p$. Thus Eq.({\ref{6d}) ties the values of $\psi_m$ at $f \gg \mu$ with those at $f \sim \mu$.

To separate the angular variables in Eq.(\ref{6d}) we present the
wave function of the bound state with projection $m=0,\pm1$ of the orbital momentum $\ell=1$ as
$$\psi_m({\bf r})=\sqrt{\frac{3}{4\pi}}r_m\varphi(r); \quad \varphi(0) \neq 0,$$
$r_m$ is the projection of the position vector ${\bf r}$. The radial part of the function is $R(r)=r\varphi(r)$. It is normalized by the condition $\int_0^{\infty}drr^2R^2(r)=1.$
In the momentum presentation
\begin{equation}
\psi_m({\bf f})= i\sqrt{\frac{3}{4\pi}}({\bf\nabla}_f)_m\varphi(f); \quad   \varphi(f)=\int d^3re^{-i{\bf f}\cdot{\bf r}}\varphi(r).
\label{9}
\end{equation}
Employing this expression and integrating by parts we write Eq.(\ref{6d}) as
$$\psi_m({\bf p})=i\sqrt{\frac{3}{4\pi}}({\bf\nabla}_p)_m\varphi(p),$$
with
\begin{equation}
\varphi(p)=-\frac{2m_e}{p^2}J^{(r)}(p) ; \quad J^{(r)}(p)=\int \frac{d^3f}{(2\pi)^3}V({\bf p}-{\bf f})\varphi(f).
\label{9at}
\end{equation}
The upper index $(r)$ reminds that  $J^{(r)}$ is connected to the radial part of the wave function.

Using Eq.({\ref{9}) and integrating by parts we write Eq.(\ref{6}) as
\begin{equation}
F^a_m=-i\sqrt{\frac{3}{4\pi}}\frac{N(\omega)}{\omega}\frac{{\bf e}\cdot {\bf p}}{m_e}({\bf\nabla}_p)_mJ^{(r)}(p),
\label{11}
\end{equation}
or
\begin{equation}
F^a_m=-i\sqrt{\frac{3}{4\pi}}\frac{N(\omega)}{\omega}\frac{{\bf e}\cdot {\bf p}}{m_e}n_mJ^{(r)'}(p); \quad {\bf n}=\frac{\bf p}{p}.
\label{12}
\end{equation}
with $J^{(r)'}(p)=\partial J^{(r)}(p)/\partial p$.

Now we include the possibility that the large recoil momentum is transferred by the photoelectron.
The corresponding amplitude is $F^b_m$, and the total amplitude is
\begin{equation}
F_m=F^a_m+F^b_m
\label{13}
\end{equation}
The photoelectron function in the field of the recoil ion satisfies the Lippmann-Schwinger equation
$\psi_{\bf p}=\psi_{\bf p}^{(0)}+G(\omega+\varepsilon_B)V\psi_{\bf p},$
with $G_b$ the propagator of free motion.
Transfer of the large momentum $p$ is expressed in asymptotics by the first iteration
$$\psi_{\bf p}^{(1)}=G(\omega+\varepsilon_B)V\psi_{\bf p}^{(0)}.$$
This can be written as
$$\psi_{\bf p}^{(1)}({\bf f})=\langle {\bf f}|G(\omega+\varepsilon_B)V|{\bf p}\rangle=g(\omega+\varepsilon_B)V({\bf p}-{\bf f}),$$
$$g(\omega+\varepsilon_B)=\frac{1}{\omega+\varepsilon_B-f^2/2m_e}.$$
We must put
$g(\omega+\varepsilon_B)=1/\omega$ in the asymptotics.

This provides
\begin{equation}
F_m^b=\frac{N(\omega)}{\omega}\int\frac{d^3f}{(2\pi)^3}V({\bf p}-{\bf f})\frac{{\bf e}\cdot {\bf f}}{m_e}\psi_m({\bf f})
\label{14}
\end{equation}
Employing Eq.(\ref{9}) and integrating by parts we find
\begin{equation}
F_m^b=F^{b(1)}_m+F^{b(2)}_m,
\label{15}
\end{equation}
with
\begin{equation}
F^{b(1)}_m=-i\sqrt{\frac{3}{4\pi}}\frac{N(\omega)}{\omega}\frac{e_m}{m_e}J^{(r)}(p),
\label{16}
\end{equation}
with $J^{(r)}$ defined by the second equality of Eq.(\ref{9at}).
The second term on the right hand side of Eq.(\ref{15})is
$$F^{b(2)}_m=i\sqrt{\frac{3}{4\pi}}\frac{N(\omega)}{\omega}n_m\cdot B'(p); \quad B(p)=\int \frac{d^3f}{(2\pi)^3}V({\bf p}-{\bf f})\frac{{\bf e}\cdot {\bf f}}{m_e}\varphi(f),$$
$B'(p)=\partial B(p)/\partial p$.
Comparing this expression with Eq.(\ref{12}) for $F^a_m$ one can see that $F^{b(2)}_m/F^a_m \sim \mu/p \ll 1$ since the integral is saturated by small
$ f \sim \mu$. This enables us to neglect $F^{b(2)}_m$, presenting $F_m=F^a_m+F^{b1}_m$.
Hence
\begin{equation}
F_m=-i\sqrt{\frac{3}{4\pi}}\frac{N(\omega)}{m_e\omega}\Big({\bf e}\cdot {\bf n}n_mpJ^{(r)'}(p)+e_mJ^{(r)}(p)\Big); \quad {\bf n}=\frac{\bf p}{p}.
\label{17h}
\end{equation}
Recall that the two terms in the parenthesis on the right hand side describe the transfer of large recoil momentum by the bound electron and by the photoelectron correspondingly.

If the logarithmic derivative of the potential is large enough, i.e. if Eq.(\ref{4}) is true, transfer of the recoil momentum by the photoelectron
can be neglected, and
\begin{equation}
F_m=-i\sqrt{\frac{3}{4\pi}}\frac{N(\omega)}{m_e\omega}{\bf e}\cdot {\bf n}n_mpJ^{(r)'}.
\label{18}
\end{equation}
If Eq.(\ref{5}) is true, both terms on the right hand side of Eq.(\ref{17h}) should be included. In this case further evaluation of the expression
 for the amplitude is possible since the integral on the right hand side of the second equality of Eq.(\ref{9at}) is saturated by small $f \sim \mu$. The integrand can be expanded in powers of $f$. Once Eq.(\ref{5}) holds,
 we find $\mu V'(p) \ll V(p)$, and we can put $V({\bf p}-{\bf f})=V({\bf p})$. Thus
\begin{equation}
J^{(r)}(p)=V(p)\varphi(r=0),
\label{19}
\end{equation}
and
\begin{equation}
F_m=-i\sqrt{\frac{3}{4\pi}}\frac{N(\omega)}{m_e\omega}\Big({\bf e}\cdot {\bf n}n_mpV'(p)+e_mV(p)\Big)\varphi(r=0); \quad {\bf n}=\frac{\bf p}{p}.
\label{20}
\end{equation}

Note that the third situation with $|pV'(p) |\ll -V(p)$ is not possible for the potentials which satisfy the condition presented by Eq.(\ref{8}).
For $p \gg \mu$ we can write $V(p)=-f(p)/p$, with $f(p)>0$ and $f'p)<0$. Calculating the derivative we find that $|pV'(p) |\geq -V(p)$

Now we can obtain the asymptotic expressions for the cross sections.
If $|pV'(p) |\gg -V(p)$ we find employing Eqs.(\ref{18}) and (\ref{20a})
\begin{equation}
\sigma(\omega)=n_e\frac{\alpha}{3\pi}\frac{p}{\omega^2}|J^{(r)'}(p)|^2;\quad p=\sqrt{2m_e\omega},
\label{219}
\end{equation}
with $n_e$ the number of electrons bound in the $p$ state. If $|pV'(p) |\sim -V(p)$, Eqs.(\ref{20}) and (\ref{20a}) provide
\begin{equation}
\sigma(\omega)=n_e\frac{\alpha}{3\pi}\frac{1}{p\omega^2}\Big(p^2V^{'2}(p)+2pV(p)V'(p)+3V^2(p)\Big)\varphi^2(r=0),
\label{229}
\end{equation}

\section{Analytical properties of potential and the asymptotic behavior of the cross section}
\subsection{Potentials with singularities at the origin}
Start with the Coulomb potential created by the point nucleus with the charge $Z$. It is $V_C(r)=-\alpha Z/r$, with a pole at $r=0$.
Its Fourier transform is
\begin{equation}
V_C(p)=\frac{-4\pi\alpha Z}{p^2}.
\label{23}
\end{equation}
In this case $pV'(p)=-2V(p)$, and Eq.(\ref{5}) is true. The large recoil momentum can be transferred by both the bound electron and the photoelectron. The cross section is described by Eq.(\ref{229}). Note that the first and second terms in parenthesis on the right hand side of Eq.(\ref{229})cancel.
 Thus only transfer of large recoil momentum by the photoelectron becomes important \footnote{Of course, this remarkable cancelation can be obtained by using the explicit expressions for photoinization cross sections of the $2p$ states in the Coulomb field obtained long ago (see, e.g. \cite{5}). However, as far as we know, it was discussed for the first time in \cite{6}.} We find
\begin{equation}
\sigma(\omega)=n_e16 \alpha\pi\frac{\alpha^2 Z^2\varphi^2(r=0)}{\omega^2 p^5}.
\label{24}
\end{equation}
The asymptotic cross section for photoionization of $np$ state can be obtained by employing the Coulomb value
 $\varphi^2(r=0)=(m\alpha Z)^5/n^5$. The expression for the cross section at $n=2$ is presented in the book \cite{5}.
 The cross section $\sigma_{np}$ drops as $\omega^{-9/2}$. Recall that the asymptotics for photoionization cross section $\sigma_{ns}$ of $s$ states
 is $\omega^{-7/2}$.  Thus the ratio $r(\omega)$ defined by Eq.(\ref{8a}) is proportional to $\omega$.

The Fourier transform of the Yukawa potential $V_{\lambda}(r)=-ge^{-\lambda r}/r$ ($g>0$) is
\begin{equation}
V_{\lambda}(p)=\frac{-4\pi g}{p^2+\lambda^2}.
\label{25}
\end{equation}
One can see that $V_{\lambda}(p)=-4\pi g/p^2$ at large $p \gg \lambda$,
and the asymptotics of the cross section in the Yukawa field coincides with that in the Coulomb field with the charge of the nucleus $Z=g/\alpha$.
The shape of the cross section is determined by Eq.(\ref{24}). The difference of the cross section values in the fields $V_C(r)$ and $V_{\lambda}(r)$
manifests itself through the difference of the values of $\varphi^2(r=0)$.

The exponential potential is
\begin{equation}
V_{exp}(r)=-V_0e^{-\lambda r}; \quad V_0>0,
\label{27}
\end{equation}
Being treated as a function of $x$ in the interval $-\infty <x < \infty$ it can be written as $V_{E}(x)=-V_0e^{-\lambda |x|}$, with a cusp at
$x=0$. We can present
$$V_{exp}(r)=-\frac{V_0}{g}\frac {\partial V_{\lambda}}{\partial \lambda},$$
and thus
\begin{equation}
V_{exp}(p)=\frac{-8\pi V_0 \lambda}{p^4}.
\label{28}
\end{equation}
In this case $pV'(p)=-4V(p)$. The asymptotics of the photoionization cross section is given by Eq.(\ref{229}) with all three terms in the parenthesis on the right hand side being important. The large recoil momentum can be transferred by both recoil electron and the photoelectron. We find
\begin{equation}
\sigma(\omega)=n_e \frac{11\cdot 64}{3} \alpha\pi\frac{V_0^2\lambda^2\varphi^2(r=0)}{\omega^2 p^{9}}.
\label{29}
\end{equation}
The cross section drops as $\omega^{-13/2}$. Note that similar analysis for photoionization of $s$ states provides $\sigma_{s} \sim \omega^{-11/2}$.
Thus as well as in previous cases $r(\omega) \sim \omega$.

In the examples presented above the shapes of the cross sections are determined directly by the shape of the potential $V(p)$.
The cross sections contain the squared wave function $\varphi^2(r=0)$ at the singular point of the potential $V(r)$.

\subsection{Potentials with singularities on the real axis}

For such potentials it is instructive to write expression for $J^{(r)'}(p)$ provided by second equality of Eq.({\ref{9at}) in the position presentation
\begin{equation}
J^{(r)}(p)=\int{d^3r}V(r)\varphi(r)e^{-i{\bf p}{\bf r}}=\frac{4\pi}{p}\int_0^{\infty}drV(r)r\varphi(r)\sin{(pr)}.
\label{31}
\end{equation}
In the asymptotics
\begin{equation}
J^{(r)'}(p)=\frac{4\pi}{p}\int_0^{\infty}drV(r)\chi(r)\cos{(pr)}; \quad \chi(r)=r^2\varphi(r).
\label{32}
 \end{equation}
Consider first the rectangular well potential for which $V(r)=-V_0<0$ for $0 \leq r \leq R$ while $V(r)=0$ for $r>R$.
It has a  singularity on the real axis at $r=R$, where the function $V(r)$ experience a jump.

The asymptotics of the Fourier transform for this potential is
\begin{equation}
V(p)=\frac{4\pi V_0R}{p^2}\cos{pR},
\label{30}
\end{equation}
and $pV'(p) \gg -V$ at $p \rightarrow \infty$. Thus the large recoil momentum is transferred by the bound state electron,
and the amplitude is determined by Eq.(\ref{18}).

To calculate the asymptotics of the integral $J^{(r)'}(p)$ we integrate by parts the integral on the right hand side of Eq.(\ref{32}).
This provides
\begin{equation}
J^{(r)'}(p)=-\frac{4\pi V_0}{p^2}\Big(\sin{(pR)}\chi(R)-\int_0^{R}dr\sin{(pr)}\chi'(r)\Big).
\label{34}
\end{equation}
The asymptotics is determined by the first term in the parenthesis on the right hand side. Further integration by parts
of the second term provides additional factors $1/p$, leading to the terms which contribute beyond the asymptotics.
Thus for the  rectangular well potential
\begin{equation}
J^{(r)'}(p)=-\frac{4\pi V_0\chi(R)\sin{(pR)}}{p^2}.
\label{35}
\end{equation}
Employing Eq.(\ref{219}) we find
\begin{equation}
\sigma(\omega)=n_e \frac{16}{3}\alpha \pi\frac{V_0^2\chi^2(R)}{\omega^2p^3}\sin^2{pR},
\label{21}
\end{equation}

In a more general case the potential $V(r)$ experience a jump on the real axis at certain $r=R$.
\begin{equation}
V(r)=V_1(r) \quad 0 \leq r \leq R; \quad V(r)=V_2(r)\quad R <r <\infty,
\label{36}
\end{equation}
with $V_1(R)\neq V_2(R)$. Similar to Eq.(\ref{32}) we write
\begin{equation}
J^{(r)'}(p)=\frac{4\pi}{p}\int_0^RdrV_1(r)\chi(r)\cos{(pr)}+\frac{4\pi}{p}\int_R^{\infty}drV_2(r)\chi(r)\cos{(pr)}.
\label{37}
\end{equation}
The function $\varphi(r)$ is continuous on the real axis. Thus integration by parts provides
\begin{equation}
J^{(r)'}(p)=-\frac{4\pi\chi(R)\sin{(pR})\delta(R)}{p^2} ;\quad \delta(R)=V_2(R)-V_1(R),
\label{38}
\end{equation}
leading to the cross section
\begin{equation}
\sigma(\omega)=n_e \frac{16}{3}\alpha \pi\frac{(\delta (R))^2\chi^2(R)}{\omega^2p^3}\sin^2{(pR)},
\label{39}
\end{equation}
Carrying out similar calculation for $s$ states
we find that
Eq.(\ref{8a}) provides the oscillating cross sections ratio
$r(\omega) \sim \cos^2{(pR)}/\sin^2{(pR)}$.

The cross section of photoionization in the field $V(r)$ which is continuous on the real axis while the derivative $V'(r)$ of the potential experience a jump at $r=R$
can be obtained in the same way. One needs two integrations by parts on the right hand side of Eq.(\ref{37}).
The function $J^{(r)'}(p)$ obtains additional factor $1/p$. The cross section drops as $\omega^{-9/2}$. The case when only higher derivatives experience jumps can be considered in similar way. We shall provide an example in Subsec.4.2.

\subsection{Potentials with singularities in the complex plane}

Consider now a simple potential
\begin{equation}
V(r)=-\frac{U_0}{\pi}\frac{a}{r^2+a^2};  \quad a>0,
\label{41}
\end{equation}
$U_0>0$ is a dimensionless constant.
Its Fourier transform is
\begin{equation}
V(p)=-\frac{2\pi U_0a}{p}\exp{(-pa)}.
\label{42}
\end{equation}
The exponential drop is due to the contributions of the poles $r=\pm ia$ in integration in the complex plane.

We can proceed further in the same way as we did for the $s$ states in \cite{3}. The integral
\begin{equation}
J^{(r)}(p)=-2\pi U_0\int\frac{d^3f}{(2\pi)^3}\frac{e^{-v({\bf f})a}}{v(\bf f)}\varphi(f); \quad v({\bf f}) =|{\bf p}-{\bf f}|= (p^2-2{\bf p}{\bf f}+f^2)^{1/2},
\label {46a}
\end{equation}
is dominated by small $f \sim 1/a \ll p$. To find the asymptotics one can put $v=p$ in the denominator of the integrand. However this can not be done in the power of the exponential factor. To obtain the $p$ dependence of the right hand side we present Eq.($\ref{46a}$)as
$$J^{(r)}(p)=-2\pi\frac{U_0}{p}I^{(r)}(p); \quad I^{(r)}(p)=\int\frac{d^3f}{(2\pi)^3}e^{-v({\bf f})a}\varphi(f),$$
and
$$J^{(r)'}(p)=-2\pi\frac{U_0}{p}I^{(r)'}(p)$$
in the asymptotics.

To find the shape of the dependence $I^{(r)}(p)$ we calculate its  derivative
with respect to $p$
$$I^{(r)'}(p)=-a\int\frac{d^3f}{(2\pi)^3}e^{-v({\bf f})a}v'_p({\bf f})\varphi(f),$$
with $v_p'=\partial v/\partial p$. One can put $v'_p=1$ with the accuracy $1/p^2$. Thus we can write a simple differential equation
\begin{equation}
I^{(r)'}(p)=-aI^{(r)}(p).
\label{46b}
\end{equation}
with the solution $I^{(r)}(p)=\kappa_1 e^{-pa}$ where $\kappa_1$ is an unknown constant.
It can be estimated as $\kappa_1 \approx \varphi(r=0)$.
Thus
\begin{equation}
J^{(r)'}(p)=2\pi U_0a\kappa_1\frac{e^{-pa}}{p}.
\label{47}
\end{equation}
The expression for the cross section
\begin{equation}
\sigma(\omega)=n_e\frac{4\pi}{3}\alpha U_0^2a^2\frac{\exp{(-2pa)}}{\omega^2p}\kappa_1^2.
\label{47a}
\end{equation}
contains the unknown constant factor which can be estimated as $\kappa_1^2 \approx \varphi^2(r=0)$.
Recall that the cross section for photoionization of $s$ states \cite{3} contains the factor $\kappa^2 \approx \psi_{1s}^2(r=0)$.

We can make numerical predictions on the cross sections ratio
at large energies $\omega_1$ and $\omega_2$
\begin{equation}
\frac{\sigma(\omega_1)}{\sigma(\omega_2)}=\frac{\omega_2^{5/2}}{ \omega_1^{5/2} }e^{-2(p_1-p_2)a}; \quad p_i=(2m_e\omega_i)^{1/2}.
\label{47b}
\end{equation}
The equation for ionization of $s$ states has the same form \cite{3}. Thus the ratio $r(\omega)$ does not depend on $\omega$.

We can proceed in similar way for the Gaussian potential
\begin{equation}
V(r)=-V_0e^{{-r^2/a^2}}; \quad V_0>0; \quad a>0,
\label{48}
\end{equation}
with the essential singularity in the complex plane.
The Fourier transform is
\begin{equation}
V(p)=-\pi^{3/2}V_0a^3e^{-p^2a^2/4}.
\label{48a}
\end{equation}
Proceeding in the same way as in previous case we find
$$J^{(r)'}(p)=\frac{\pi^{3/2}V_0a^5\kappa_1}{2}pe^{-p^2a^2/4},$$
and
\begin{equation}
\sigma=n_e\frac{\pi^2\alpha}{2}V_0^2a^{10}\frac{m_ep}{\omega}e^{-p^2a^2}\kappa_1^2,
\label{48b}
\end{equation}
with $\kappa_1^2 \approx \varphi^2(r=0)$.
For the cross sections ratio at large values of the photon energies $\omega_{1,2}$
\begin{equation}
\frac{\sigma(\omega_1)}{\sigma(\omega_2)}=\frac{\omega_2^{1/2}}{ \omega_1^{1/2} }e^{-m_ea^2(\omega_1-\omega_2)}.
\label{47c}
\end{equation}

Comparing with the results for photoionization of $s$ states one can see that $r(\omega) \sim 1/\omega$. The cross section for photoinization of $p$ states drops slower than that for $s$ states while the photon energy increases,.

\section{Photoionization of fullerenes}

Recall that fullerene is a bound system of several dozens of the carbon atoms. The number of the atoms is usually denoted by $N$ and the fullerene is $C_N$. The $2N$ $1s$ electrons are bound to their nuclei while the other $4N$ ones are collectivized. Here we analyze the case when the fullerenes have approximately spherical shape. The fullerene $C_{60}$ is the most studied one. In such fullerenes the electrons are often assumed to move in the field described by a spherical potential $V(r)$.  The electrons are located in the layer between the spheres of radii $R-\Delta/2$ and
$R+\Delta/2$. Characteristic values of the width of the layer $\Delta$ are of the order $r_0=1 a.u.$ while the empirical values of the radii are about $R=6r_0 =6 a.u.$.  The field $V(r)$ reaches its largest values in the interval
$ R-\Delta/2 \leq r \leq R+\Delta/2$.

\subsection{The "bubble" potentials}
In the Dirac bubble potential \cite{7} the width of the layer $\Delta=0$, and
\begin{equation}
V(r)= -U_0\delta(R-r);\quad U_0>0,
\label{49}
\end{equation}
where $U_0$ is a dimensionless constant.
The Fourier transform of this potential
\begin{equation}
V(p)=-4\pi U_0R\frac{\sin{(pR)}}{p},
\label{50}
\end{equation}
does not satisfy the condition expressed by Eq. ({\ref{8}).
Hence all momenta $f \la p$ contribute to the integral for the function $J^{(r)}(p)$ --see Eq.(\ref{9at}).

Our result for the contribution $F^a_m$ to the amplitude describing transfer of large recoil momentum by the bound electron presented by Eq.(\ref{12})
is true. Our analysis of the contribution $F^b_m$ describing transfer of large recoil momentum by the photoelectron does not work in this case since it employed the smallness of the momentum $f \ll p$. However another approach based on the partial waves expansion \cite{8} demonstrates that the photoelectron can be described by the plane wave in the asymptotics. Thus the photoionization cross section is given by Eq.(\ref{219}).
Since
\begin{equation}
J^{(r)}(p)=-4\pi U_0R\frac{\sin{(pR)}}{p}\varphi(R),
\label{51}
\end{equation}
the cross section is
\begin{equation}
\sigma=n_e\frac{16}{3}\alpha \pi U_0^2\chi^2(R)\frac{\cos^2{(pR)}}{\omega^2p},
\label{52z}
\end{equation}
dropping as $\omega^{-5/2}$. This leads to the oscillating photoionization cross sections ratio
$r(\omega) \sim \sin^2{(pR)/\cos^2{(pR)}}$.

The Dirac bubble potential can bound one electron with orbital momenta $\ell=0,1,2$ if $R=6r_0$, $\Delta=1 a.u.$ \cite{7}.
Thus it is good for the description of photoionization of extra electron in the ion $C_{N}^{-}$ with $n_e=1$ in Eq.(\ref{52z}).
For photoionization of the fullerene $C_{N}$ it is more reasonable to consider the valence electrons to be moving in the Lorentz bubble potential
\begin{equation}
V(r)= -\frac{U_0}{\pi}\frac{a}{(r-R)^2+a^2},
\label{49a}
\end{equation}
where $U_0>0$ is a dimensionless constant, $a \ll R$. At $a \rightarrow 0$ this is just the Dirac bubble potential given by Eq.(\ref{49}).
We can proceed in the same way as we did in \cite{2} for photoionization of $s$ states. The Fourier transform of the potential determined by Eq(\ref{49a}) is  $V(p)=V_A(p)+V_B(p)$, where
$$V_A(p)=-4\pi\frac{U_0R}{p}e^{-pa}\sin{(pR)},$$
and
$$V_B(p)=\frac{16U_0a}{p(pR)^3},$$
coming from the regions $|r-R| \ll R$ and $r \sim 1/p$ correspondingly. The contribution $V_B$ becomes important in the region of the photoelectron energies of the order of several keV where the cross section becomes unobservably small. Thus we can put $V(p)=V_A(p)$ in the region of the photon energies which are of physical interest. Now the condition expressed by Eq.(\ref{4}) is true and the cross section is given by Eq.(\ref{219}).
We find
\begin{equation}
\sigma=n_e\frac{16}{3}\alpha \pi U_0^2\chi^2(R)\frac{e^{-2pa}\cos^2{(pR)}}{\omega^2p}.
\label{52a}
\end{equation}
As well as in the case os $s$ states this is just the cross section for photoionization in the field of the Dirac bubble potential multiplied
by the exponential factor $e^{-2pa}$.

\subsection{Jellium model}

Now the positive charge of the core is assumed to be uniformly distributed inside the fullerene layer \cite{9}.
Its field can be approximated by the potential for which $V(r)=V_1(r)$ at $0\leq r <R_1$, $V(r)=V_2(r)$ at $R_1\leq r  \leq R_2$  and $V(r)=V_3(r)$ at $ r > R_2$, where
\begin{equation}
V_1(r)\ =-U_0\frac{3}{2}\frac{R_2^2-R_1^2}{R_2^3-R_1^3} ; \quad V_2(r)=-\frac {U_0}{2(R_2^3-R_1^3)}\Big(3R_2^2-r^2(1+\frac{2R_1^3}{r^3})\Big);
\label{53}
\end{equation}
$$V_3(r)=-\frac{U_0}{r}; \quad U_0>0.$$
Here $U_0$ is the dimensionless constant. This potential is continuous at the real axis. The same refers to its derivatives. The second derivatives
$V^{(2)}(r)$ experience jumps at $r=R_{1,2}$.

In our case Eq.(35) takes the form
$$J^{(r)}(p)=\frac{4\pi}{p}\Big(\int_0^{R_1}drV_1(r)r\varphi(r)\sin{(pr)}+$$
$$\int_{R_1}^{R_2}dr V_2(r)r\varphi(r)\sin{(pr)}+\int_{R_2}^{\infty}drV_3(r)r\varphi(r)\sin{(pr)}\Big).$$
We need three integrations by parts to obtain the asymptotics
\begin{equation}
J^{(r)}(p)=\frac{4\pi}{p^4}(\zeta(R_1)+\zeta(R_2)); \quad \zeta(R_i)=-R_i\varphi(R_i)\cos{(pR_i)}\delta(R_i) ; \quad i=1,2;
\label{54}
 \end{equation}
with $\delta(R_1)=V^{''}_2(R_1)-V^{''}_1(R_1)$,  $\delta (R_2)=V^{''}_3(R_2)-V^{''}_2(R_2)$-the jumps of the second derivatives $V''(r)$ of the potential $V(r)$.
In the limit $\Delta \ll R$  they are
 $$\delta(R_1)=\frac{U_0}{R^2\Delta}; \quad \delta(R_2)=-\frac{ U_0}{R^2\Delta}.$$

The asymptotics of the Fourier tramsform of the jellium model potential can be obtained by putting $\varphi(r)=1$ in expressions for
$J^{(r)}(p)$ provided by Eqs. (35) and (63)
$$V(p)=-\frac{-8\pi U_0}{\Delta R}\frac{\sin{(p\Delta/2)}\sin{(pR)}}{p^4}.$$
One can see that Eq.(\ref{4}) is true. The large recoil momentum is transferred by the bound electron.

The asymptotics of cross section is provided by Eq.(27). In the lowest order of expansion in powers of $\Delta/R$
$$J^{(r)'}(p)=\frac{4\pi}{p^4}\frac{U_0}{\Delta}(\varphi(R_1)\sin{(pR_1)}-\varphi(R_2)\sin{(pR_2)}).$$
The cross section is
\begin{equation}
\sigma=n_e\frac{16}{3}\alpha\pi\frac{U_0^2}{\Delta^2}\frac{\Big(\varphi(R_1)\sin{(pR_1)}-\varphi(R_2)\sin{(pR_2)}\Big)^2}{\omega^2p^7}.
\label{52}
\end{equation}
It drops as $\omega^{-11/2}$. Employing results for photoionization of $s$ states \cite{2} we find the oscillating ratio $r(\omega)$ defined by Eq.(\ref{8a}).

\section{Summary}

We demonstrated that the high energy nonrelativistic  asymptotics of photoionization  cross section if the single--particle states with $\ell=1$
bound by a central field $V(r)$ can be obtained without solving the wave equations for the electrons. The asymptotics can be expressed in terms of the Fourier transform $V(p)$ of the potential and its first derivative $V'(p)$. The large recoil momentum can be transferred to the nucleus either by
the bound electron or by the photoelectron. The role of the two mechanisms depends on the relation between $V(p)$ and $V'(p)$. The latter depends on the form ("gauge") of the electron-photon interaction. We work in the velocity gauge. If the derivative $V'(p)$ is large enough and Eq.(4) is true, the recoil momentum is transferred by the bound electron. If Eq.(5) is true, both mechanisms should be included.

On the other hand relation between $V(p)$ and $V'(p)$ depends on the analytical properties of the potential $V(r)$. For the potentials with a singularity at $r=0$ Eq.(5) is true and both mechanisms should be included. In the important case of the potentials with Coulomb behavior at $r=0$
with $pV'(p)/V(p)=-2$ the squared amplitude describing transfer of recoil momentum  by the bound electron is totally canceled by the term describing the interference of the two mechanisms in the expression for the cross section. Thus the latter is determined by transfer of large momentum by the photoelectron. The the case of exponential potential with  the cusp at $r=0$ and $pV'(p)/V(p)=-4$ both mechanisms are at work. In both examples
the cross section of ionization of $p$ states drops faster than that of $s$ states and the ratio $r(\omega)$ determined by Eq.(\ref{8a}) exhibits a linear rise with $\omega$.

For the potentials $V(r)$ with discontinuity on the real axis we presented the cross section in terms of the jump of the potential $V(r)$ at
the discontinuity point. The large recoil momentum is transferred by the bound electron. The same approach can be applied for potentials
which are continuous on the real axis while the its derivatives experience jumps. In these cases the ratio $r(\omega)$ oscillates.

If the binding field is approximated by an analytical function $V(r)$ with poles in the complex plane, the cross section exhibits exponential drop.
The shape of the energy dependence of the cross section is determined by the derivative $V'(p)$. The ratio $r(\omega)$ does not depend on the photon energy. For the  Gaussian potential with essential singularity in the complex plane the cross section exhibits the Gaussian-type drop. The cross section of photoionization of $p$ states drops slower than that of $s$ states, and $r(\omega) \sim 1/\omega$.

The results can be applied in the studies of photoionization of fullerenes.  Here the model potentials for the description of interaction between the valence electrons and the core are often used. We provided two examples, calculating the asymptotics of the photoionization cross section in the Dirac bubble model and in the jellium model. In the former case the potential experiences the infinite jump on the rear axis. In the latter case the potential is continuous, while the second derivatives experience jumps. In both case the ratio $r(\omega)$ exhibits oscillating behavior.

\end{document}